\documentstyle[11pt,epsf]{article}
\begin{document}
\setcounter{page}{1}
\pagestyle{plain}
\setcounter{equation}{0}
%
%
\ \\[13mm]
\begin{center}
	{\bf 	MATRIX PRODUCT GROUND STATES FOR EXCLUSION PROCESSES\\[1mm]
                WITH PARALLEL DYNAMICS} \\[25mm]
\end{center}
\begin{center}
\normalsize
	Haye Hinrichsen\footnote{
        e-mail: \tt fehaye@wicc.weizmann.ac.il}\\[13mm]
	{\it Department of Physics of Complex Systems\\
	Weizmann Institute of Science\\ Rehovot 76100, Israel}
\end{center}
\vspace{30mm}
{\bf Abstract:}
We show in the example of a one-dimensional asymmetric exclusion process that
stationary states of models with parallel dynamics may be written
in a matrix product form. The corresponding algebra is quadratic and
involves three different matrices. Using this formalism we
prove previous conjectures for the equal-time correlation
functions of the model.
\\[25mm]
\\[3mm]
\rule{6.5cm}{0.2mm}
\begin{flushleft}
\parbox[t]{3.5cm}{\bf Key words:}
\parbox[t]{12.5cm}{reaction-diffusion models,
                   asymmetric exclusion process,\\
                   matrix product ground states,
		   parallel dynamics}
\\[2mm]
\parbox[t]{3.5cm}{\bf PACS numbers:} 02.50.Ey, 05.60.+w, 05.70.Ln
\parbox[t]{12.5cm}{}
\end{flushleft}
\normalsize
\thispagestyle{empty}
\mbox{}
\pagestyle{plain}
%
%
%
%
\newpage
\setcounter{page}{1}
\setcounter{equation}{0}
During the last few years the study of one-dimensional reaction-diffusion
models has been of increasing interest. These models describe dynamical
processes far away from thermal equilibrium so that in general their stationary
probability distribution cannot be derived from an energy function.
Therefore different techniques are needed in order to determine the stationary
properties. An exact method which turned out to be very
successful is the so-called matrix product formalism
\cite{HakimNadal}-\cite{OurLastPaper}.
This formalism can be seen as a generalization of stationary
states with a product measure in which products of numbers are replaced
by products of non-commutative algebraic objects. By representing these objects
in terms of matrices, the stationary state and all equal time correlation
functions can be derived exactly. Up to now this technique has been applied
mostly to systems with sequential dynamics (continuous time evolution)
where the stationarity of the state is related to an additive
cancelation mechanism from site to site. However, many systems,
for example traffic models \cite{TrafficModels}, are defined by
parallel dynamics rather than sequential updates. Therefore it is of
interest to find applications of the matrix product technique to
systems with parallel dynamics. The present work discusses this problem
in the example of a one-dimensional asymmetric exclusion process with
parallel updates \cite{Gunter}. A modified matrix product formalism
is presented in which a new multiplicative cancelation
mechanism plays an essential role. The corresponding matrix algebra is
derived and finite-dimensional representations are given. This allows
to prove exact results for the equal-time correlation functions which were
already given as conjectures in Ref. \cite{Gunter}.

Let us first recall the matrix product formalism for
one-dimensional reaction-diffusion models with sequential dynamics
and open boundary conditions.
A two-state model with $L$ sites is said to have a matrix product ground
state if the stationary probability
distribution $P_0(\tau_1,\tau_2,\ldots,\tau_L)$ can be written as
\begin{eqnarray}
\label{UsualMatrixAnsatz}
P_0(\tau_1,\tau_2,\ldots,\tau_L) \;&=&\;{Z}^{-1}\;
\langle W| \,
\prod_{j=1}^L (\tau_j D \,+\, (1-\tau_j)E\,)\,
|V\rangle
\end{eqnarray}
where $\tau_j \in \{0,1\}$ is the occupation number at site $j$.
$E$ and $D$ are square matrices acting in an auxiliary space
which may be either finite or infinite dimensional. The probabilities
are the expectation values $\langle W|\ldots|V\rangle$ of the matrix
products normalized by the constant $Z=\langle W|(D+E)^L |V\rangle$.
Formally we may rewrite Eq. (\ref{UsualMatrixAnsatz}) as a tensor product
\begin{equation}
\label{UsualGroundState}
|P_0\rangle \;=\;{Z}^{-1}\; \langle W| \,
\left( \hspace{-1.5mm} \begin{array}{c} E \\[-1mm]
 D \end{array} \hspace{-1.5mm}  \right) ^ {\otimes L}
|V\rangle\,\,
\end{equation}
where the vector $|P_0\rangle$ represents the stationary
probability distribution in configuration space.
The matrix product representation is a powerful tool since it allows
various physical quantities like the particle density
\begin{equation}
\label{ParticleDensity}
\langle \tau_j \rangle_L \;=\;
\frac{\langle W | \, C^{j-1} D C^{L-j} \, | V \rangle}
     {\langle W | \, C^L \, | V \rangle}
\hspace{15mm}
(C=D+E)\,
\end{equation}
to be computed directly. Higher correlation functions
are given by similar expressions in which $C$ plays
the role of a transfer matrix. However, a special mechanism
is needed in order to ensure that the state in Eq. (\ref{UsualGroundState})
is indeed a stationary one. For models with sequential dynamics
this mechanism amounts in an additive cancelation from site
to site: Assume that the time evolution of the
system is described by a master equation
$\frac{d}{dt} |P\rangle = -H|P\rangle$ with a time evolution operator
$H=\sum_{j=1}^{L-1} h_{j,j+1} + h^{(L)}_1 + h^{(R)}_L$,
where $h_{j,j+1}$ is a $4 \times 4$ interaction matrix and
$h^{(L)}$ and $h^{(R)}$ are $2 \times 2$ matrices for particle
input and output at the ends of the chain. Then the matrices
$E$ and $D$ have to be chosen such that the application of the
interaction matrix $h_{j,j+1}$ to a pair of sites results in a
local divergence-like term on the right hand side
\begin{equation}
\label{OldAnsatzBulk}
h \, \left[ \,
\left( \hspace{-1.5mm} \begin{array}{c} E \\[-1mm]
 D \end{array} \hspace{-1.5mm} \right) \otimes
\left( \hspace{-1.5mm} \begin{array}{c} E \\[-1mm]
 D \end{array} \hspace{-1.5mm} \right)
\, \right] \;\;=\;\;
\left( \hspace{-1.5mm} \begin{array}{c} \hat{E} \\[-1mm]
 \hat{D} \end{array} \hspace{-1.5mm} \right) \otimes
\left( \hspace{-1.5mm} \begin{array}{c} E \\[-1mm]
 D \end{array} \hspace{-1.5mm} \right) -
\left( \hspace{-1.5mm} \begin{array}{c} E \\[-1mm]
 D \end{array} \hspace{-1.5mm} \right) \otimes
\left( \hspace{-1.5mm} \begin{array}{c} \hat{E} \\[-1mm]
 \hat{D} \end{array} \hspace{-1.5mm} \right)
\;\;,
\end{equation}
where $\hat{E}$ and $\hat{D}$ are again matrices in the auxiliary space.
By summing up the two-particle interactions, all these contributions
cancel in the bulk of the chain. The remaining terms at the boundaries
have to be canceled by the surface fields for particle input and output:
\begin{equation}
\label{NewSurfaceTerms}
<W| \, h^{(L)} \left( \hspace{-1.5mm} \begin{array}{c} E \\[-1mm]
 D \end{array} \hspace{-1.5mm} \right) =
-<W| \, \left( \hspace{-1.5mm} \begin{array}{c} \hat{E} \\[-1mm]
 \hat{D} \end{array} \hspace{-1.5mm} \right)\,,
\hspace{15mm}
h^{(R)} \left( \hspace{-1.5mm} \begin{array}{c} E \\[-1mm]
 D \end{array} \hspace{-1.5mm} \right) \, |V> =
\left( \hspace{-1.5mm} \begin{array}{c} \hat{E} \\[-1mm]
 \hat{D} \end{array} \hspace{-1.5mm} \right)\, |V> \,.
\end{equation}
The simplest case $\hat{E}=\hat{D}=0$ and its generalization
to spin one problems has been considered in Ref. \cite{SimplestType}.
Another system which has been investigated in detail
is the (asymmetric) exclusion process where
$\hat{E}=-\hat{D}=1$ \cite{DiffusionType}.
In both cases one is led to an quadratic
algebra of two objects $E$ and $D$ \cite{QuadraticAlgebra}.
In a similar way matrix product ground states were found
for particular three-state models \cite{ThreeState}.
Also excited states can be described with a matrix ansatz
\cite{Stinchcombe} where $\hat{E}=-\hat{D}$ has to be chosen as a
time-dependent matrix leading to a quadratic algebra of three
different objects. Taking $\hat{E}$ and $\hat{D}$ as independent matrices,
it was also possible to find the stationary state
of particular models with particle reactions \cite{OurLastPaper}.

So far, the interest has been focused mainly on stochastic models with
continuous time evolution. However, similar techniques can be used
for systems with parallel dynamics. A first
example of this type was given in Ref. \cite{HakimNadal} where the
transfer matrix for a deterministic model of directed animals on
a strip was investigated. It is the aim of the present work to point out
that there could be a broad spectrum of applications to reaction
diffusion models with parallel dynamics.
For this purpose we consider a one-dimensional
asymmetric exclusion process with parallel updates which was originally
introduced by G. M. Sch\"utz in Ref. \cite{Gunter}. In this model
particles move on a one-dimensional lattice with $L=2N$ sites
and open boundaries. The bulk dynamics is deterministic
and consists of two half time steps. In the first half time step
particles at odd positions move one step to the right provided
that the neighboured site to the right is empty. Then in the second
half time step the particles at even positions move to the right in
the same way. In addition particles are injected (removed) stochastically
with rate $\alpha$ ($\beta$) at the left (right) boundary:

%
%
\vskip-5mm
\epsfysize=65mm
\centerline{\epsfbox{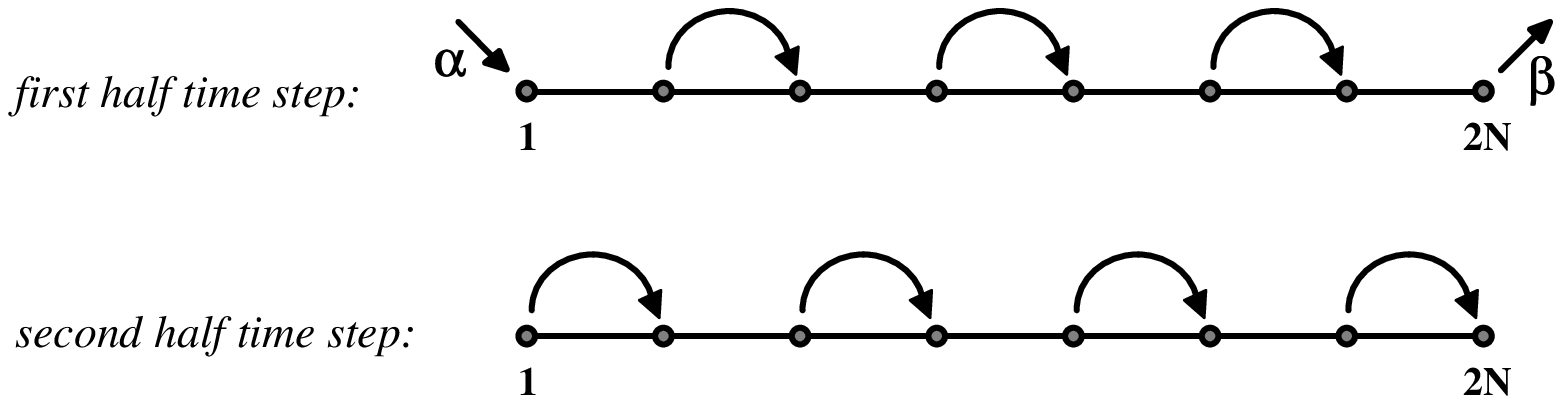}}
\vskip-10mm

\noindent
The corresponding transfer matrix therefore consists
of two factors $T=T_2 T_1$
\begin{eqnarray}
\label{T1}
T_1 &=& {\cal L} \otimes {\cal T} \otimes \ldots \otimes
        {\cal T} \otimes {\cal R} \,\;=\;\,
{\cal L} \otimes {\cal T}^{\otimes (N-1)} \otimes {\cal R}\\
\label{T2}
T_2 &=& \hspace{3.8mm} {\cal T} \otimes {\cal T} \otimes \ldots \otimes
        {\cal T} \hspace{3.8mm} \,\;=\;\,  {\cal T}^{\otimes N} \nonumber
\end{eqnarray}
where ${\cal T}$, ${\cal L}$ and ${\cal R}$ are the matrices for hopping,
particle input and output:
\begin{equation}
\label{Interaction}
{\cal T} \;=\; \left(
\begin{array}{cccc}
1 & 0 & 0 & 0 \\
0 & 1 & 1 & 0 \\
0 & 0 & 0 & 0 \\
0 & 0 & 0 & 1
\end{array} \right)\,,
\hspace{10mm}
{\cal L} \;=\; \left(
\begin{array}{cc}
1-\alpha & 0 \\
\alpha & 1
\end{array} \right)\,,
\hspace{10mm}
{\cal R} \;=\; \left(
\begin{array}{cc}
1 & \beta \\
0 & 1-\beta
\end{array} \right) \,.
\end{equation}
The phase diagram of this model shows two phases.
For $\alpha<\beta$ the system is in a low-density phase
with an average particle density $\rho=\alpha/2 < 1/2$ whereas in
the high-density phase $\alpha>\beta$ one has $\rho=1-\beta/2 > 1/2$.
The total current in the thermodynamic limit is given by
$j=min(\alpha,\beta)$. The physical behaviour is closely related
to that of asymmetric exclusion models with continuous
time evolution \cite{DiffusionType} (there is an additional
phase with maximal density in the latter case). It plays a role
in traffic models \cite{TrafficModels} as well as in polymer physics
\cite{PolymerPhysics}.
Related models with deterministic dynamics can be found in Ref.
\cite{DeterministicModels} and the influence of defects has been studied
in Ref.~\cite{WithDefects}.
%
%
%

As we are going to show below, the stationary state of the
exclusion model (\ref{T1})-(\ref{Interaction}) can be written as a
matrix product state with alternating pairs of matrices
$(E,D)$ and  $(\hat{E},\hat{D})$ so that the probability
to find the system in the configuration
$(\tau_1, \tau_2 \ldots \tau_{2N})$ is given by
\begin{equation}
\label{MatrixAnsatz}
P_0(\tau_1,\tau_2,\ldots,\tau_{2N}) \;=\;
Z^{-1}\,\langle W| \,
\prod_{i=1}^N \,  \left[\,\Bigl(
\tau_{2i-1} \hat{D} \,+\, (1-\tau_{2i-1})\hat{E}\,\Bigr)\,
\Bigl(\tau_{2i} D \,+\, (1-\tau_{2i})E\,\Bigr)\, \right]
|V\rangle \,.
\end{equation}
As in Eq. (\ref{UsualGroundState}), we may also write
%
%
\def\edmatrix{\Bigl(\hspace{-1.5mm}
    \begin{array}{c} E \\[-1mm] D
    \end{array}\hspace{-1.5mm} \Bigr)}
\def\abmatrix{\Bigl(\hspace{-1.5mm}
    \begin{array}{c} \hat{E} \\[-1mm] \hat{D}
    \end{array}\hspace{-1.5mm} \Bigr)}
\begin{eqnarray}
\label{GroundState}
|P_0\rangle &=& Z^{-1}\,
	   \langle W| \, \abmatrix \otimes \edmatrix \otimes
           \abmatrix \otimes \edmatrix \otimes \ldots \otimes
           \abmatrix \otimes \edmatrix \, |V\rangle \\[2mm]\nonumber
&=& Z^{-1}\, \langle W| \, \left[\abmatrix \otimes \edmatrix
              \right]^{\otimes N}  |V\rangle \,
\end{eqnarray}
where $Z=\langle W| \Big( (\hat{E}+\hat{D}) \otimes (E+D) \Big)
^{\otimes N} |V\rangle$.
It is obvious that in this case
the basic mechanism leading to a stationary state
has to be different from the usual one for continuous time evolution
operators. Instead of the additive cancelation from site to site
we now need a multiplicative mechanism
suitable for stationary states states of parallel
transfer matrices $T|P_0\rangle=|P_0\rangle$. In the case
of the above exclusion model this mechanism turns out to be very
simple. Let us assume that in each time step the two pairs of
matrices $(E,D)$ and $(\hat{E},\hat{D})$ are exchanged:
\begin{equation}
\label{BasicMechanism}
T_1|P_0\rangle \;=\;
\langle W| \, \left[\edmatrix \otimes \abmatrix
\right]^{\otimes N}  |V\rangle \,,
\hspace{15mm}
T|P_0\rangle \;=\; T_2T_1|P_0\rangle \;=\; |P_0\rangle
\end{equation}
This exchange mechanism can be realized by
\begin{equation}
\label{Equations}
{\cal T} \, \left[ \edmatrix \otimes \abmatrix \right] \;=\;
\abmatrix \otimes \edmatrix  \,,
\end{equation}
$$
\langle W| {\cal L} \abmatrix \;=\; \langle W| \edmatrix\,,
\hspace{15mm}
{\cal R} \edmatrix |V\rangle\;=\; \abmatrix|V\rangle \,,
$$
which is equivalent to the algebra
\begin{eqnarray}
\label{RawBulkAlgebra}
[E,\hat{E}] &=& [D,\hat{D}] \;=\; 0 \nonumber \\
E\hat{D} &=& [\hat{E},D] \\
\hat{D}E &=& 0 \nonumber
\end{eqnarray}
and the boundary conditions
\begin{equation}
\label{RawBoundaryConditions}
\begin{array}{c}
   \langle W| \hat{E} (1-\alpha) \;=\; \langle W| E  \\[1mm]
   \langle W| (\alpha \hat{E}+\hat{D}) \;=\; \langle W| D \end{array}
\hspace{20mm}
\begin{array}{c}
   (1-\beta)D|V\rangle \;=\; \hat{D}|V\rangle \\[1mm]
   (E+\beta D)|V\rangle \;=\; \hat{E}|V\rangle \end{array}\,.
\end{equation}
The commutation relations (\ref{RawBulkAlgebra})
involve four different matrices. However, only three
of them are independent since the matrix product in
Eq. (\ref{GroundState}) is invariant
under the transformation
\begin{equation}
E \rightarrow U^{-1}E\,,
\hspace{10mm}
D \rightarrow U^{-1}D\,,
\hspace{10mm}
\hat{E} \rightarrow \hat{E}U\,,
\hspace{10mm}
\hat{D} \rightarrow \hat{D}U\,.
\hspace{10mm}\end{equation}
Because of $[E+D,\hat{E}+\hat{D}]=0$ it is possible to choose
a basis in which both operators $E+D$ and $\hat{E}+\hat{D}$ are diagonal.
Taking now $U=(E+D)^{1/2}(\hat{E}+\hat{D})^{-1/2}$
both operators become identical so that we may add the relation
\begin{equation}
\label{AdditionalRelation}
C \;=\; E+D \;=\; \hat{E}+\hat{D}\,.
\end{equation}
Eliminating $\hat{E}$ and $E$ we therefore obtain
a quadratic algebra of three independent objects defined
by three bulk equations
\begin{equation}
\label{BulkAlgebra}
\hat{D}C \;=\; \hat{D}D \;=\; D\hat{D}\,, \hspace{15mm}
[D-\hat{D},C] \;=\; 0
\end{equation}
and two boundary relations
\begin{equation}
\label{BoundaryConditions}
\langle W| \,  \Bigl(D-\alpha C-(1-\alpha)\hat{D} \Bigr) \;=\; 0 \,,
\hspace{15mm}
\Bigl((1-\beta)D - \hat{D}\Bigr)\, |V\rangle \;=\; 0\,.
\end{equation}
%
%
Matrix product states based on quadratic algebras with three
objects were first studied in Ref.~\cite{Stinchcombe}.
A detailed analysis of algebras with more than two objects
and their representations will be given in Ref. \cite{VladimirNew}.

In order to check the consistency of the algebra
(\ref{AdditionalRelation})-(\ref{BulkAlgebra}) let us show
that the expectation value of any sequence of operators is given
uniquely on an abstract level. For only two operators it can be verified
by hand that
\begin{eqnarray}
\langle W|\hat{D}C|V\rangle \;=\; \langle W|\hat{D}D|V\rangle
&=& \frac{\alpha^2(1-\beta)}
         {(\alpha^2+\alpha\beta)(1-\beta)+\beta^2}
\,\langle W|CC|V\rangle\\
\langle W|CD|V\rangle
&=& \frac{\alpha^2(1-\beta)+\alpha\beta}
         {(\alpha^2+\alpha\beta)(1-\beta)+\beta^2}
\,\langle W|CC|V\rangle
\end{eqnarray}
so that any expectation value of two operators is a given number
times $Z=\langle W|CC|V\rangle$.
In order to check the consistency of the algebra for products of
arbitrary length, it is more convenient to use a different
basis of operators which is defined by the invertible transformation
\begin{eqnarray}
X &=& \frac{1}{\alpha\beta}\Bigl(D-\alpha C+(\alpha-1)\hat{D}\Bigr)\nonumber\\
Y &=& \frac{1}{\alpha\beta}\Bigl((1-\beta)D-\hat{D}\Bigr) \\
S &=& \frac{1}{\alpha\beta}(D-\hat{D})\nonumber\,.
\end{eqnarray}
In this basis, the bulk algebra (\ref{BulkAlgebra}) reads
\begin{equation}
\label{XYSBulk}
[X,S] \;=\; [Y,S] \;=\; 0 \,, \hspace{15mm}
YX \;=\; (1-\alpha)SY + (1-\beta)XS - (1-\alpha)(1-\beta)S^2
\end{equation}
and the boundary relations (\ref{BoundaryConditions}) become
particularly simple:
\begin{equation}
\label{XYSBoundary}
\langle W| X \;=\; 0 \,, \hspace{15mm}
Y |V\rangle \;=\; 0\,.
\end{equation}
As can be seen easily, the application of the bulk relations
(\ref{XYSBulk}) allows every product of $2N$ matrices
$X$, $Y$ and $S$ to be ordered as a
linear combination of terms like $X^n\,S^{2N-n-m}\,Y^m$. Since the only
nonzero expectation values of these terms is $\langle W|S^{2N}|V\rangle$,
the expectation value of any product of $2N$ matrices
is a well-defined number times $\langle W|S^{2N}|V\rangle$. The
actual value of $\langle W|S^{2N}|V\rangle$ is irrelevant since it is
canceled by the normalization constant $Z=\langle W| C^{2N} |V \rangle$.
Thus it is obvious that the algebra (\ref{XYSBulk})-(\ref{XYSBoundary})
determines the ground state $|P_0\rangle$ {\it uniquely} on an abstract level.
We should emphasize that the mathematical structure of this algebra is
different from that for exclusion models with continuous time
evolution where one has linear terms in the bulk algebra
(e.g. $DE=D+E$). Whereas in the latter case
any expectation value can be reduced to
the empty bracket $\langle W|V \rangle$, the algebra (\ref{BulkAlgebra})
does not allow to reduce the number of factors in a matrix product. Instead
of this we have shown that by means of the algebraic rules the expectation
values of all words with the same number of factors are linearly dependent.
%
%
%

%
The algebra (\ref{BulkAlgebra})-(\ref{BoundaryConditions}) can be
represented by two-dimensional matrices. For $\alpha \neq \beta$
a representation in which $C$ is diagonal is given by
\begin{equation}
\label{Rep11}
C_1 =
\left( \begin{array}{cccc}
\alpha & 0 \\
0 & \beta
\end{array} \right)
\hspace{15mm}
D_1=
\left( \begin{array}{cccc}
\alpha & 0  \\
-\alpha\beta & \alpha\beta
\end{array} \right)
\hspace{15mm}
\hat{D}_1=
\left( \begin{array}{cccc}
\alpha(1-\beta) & 0 \\
-\alpha\beta & 0
\end{array} \right)
\end{equation}
\begin{equation}
\label{Rep12}
\langle W_1| \;=\; (\alpha,\; 1-\alpha)\,,
\hspace{15mm}
|V_1\rangle \;=\;
\left( \hspace{-1.5mm} \begin{array}{c}
1-\beta \\ -\beta
\end{array} \hspace{-1.5mm} \right)
\end{equation}
The normalization constant in this representation can be computed easily and
reads
\begin{equation}
\label{Norm1}
Z_{1} \;=\; (1-\beta)\,\alpha^{2N+1} \,-\, (1-\alpha)\,\beta^{2N+1}\,.
\end{equation}
As already mentioned, the matrix $C$ acts like a transfer matrix
between the points of the correlation functions. Therefore the
length scales to be expected are essentially given by the quotients of the
eigenvalues of $C$. Thus in the present case the correlation functions
involve only a single length scale, namely $\log (\alpha/\beta)$.
This length scale diverges at the phase transition line
$\alpha=\beta$ where the constant $Z_1$ vanishes so that the above
representation becomes singular. It turns out that in this case
the operator $C$ cannot be diagonalized so that one has to use a
different representation where $C$ has a Jordan normal form:
\begin{equation}
C_2 =
\left( \begin{array}{cccc}
1 & 1 \\
0 & 1
\end{array} \right)
\hspace{10mm}
D_2 =
\left( \begin{array}{cccc}
\alpha & 1  \\
0 & 1
\end{array} \right)
\hspace{10mm}
\hat{D}_2 =
\left( \begin{array}{cccc}
0 & 1  \\
0 & 1-\alpha
\end{array} \right)
\end{equation}
\begin{equation}
\langle W_2| \;=\; (1,\; 0)\,,
\hspace{15mm}
|V_2\rangle \;=\;
\left( \hspace{-1.5mm} \begin{array}{c}
1 \\ 1-\alpha
\end{array} \hspace{-1.5mm} \right)
\end{equation}
Because of
\begin{equation}
\label{JordanNormalPower}
C_2^k \;=\; \left( \begin{array}{cccc}
1 & k \\
0 & 1
\end{array} \right)
\end{equation}
the normalization constant $Z$ is now linear in the system size:
\begin{equation}
\label{Norm2}
Z_2 \;=\; 1+2N(1-\alpha)\,.
\end{equation}
%
%

%
Using the matrix product formalism it is now easy to derive
explicit expressions for the equal-time correlation functions.
Following the ideas of Ref. \cite{Gunter}, we first compute
the $n$-point functions of the operators
\begin{equation}
\label{EtaDef}
\eta_{2j} \;=\; \frac{\tau_{2j}-\alpha}{1-\alpha}
\hspace{20mm}
\eta_{2j-1} \;=\; \frac{\tau_{2j-1}}{1-\beta}
\end{equation}
Denoting the corresponding matrices by
\begin{equation}
F_j \;=\; \left\{ \begin{array}{ll}\
(D-\alpha C)/(1-\alpha) & \mbox{if $j$ even} \\
\hat{D}/(1-\beta) & \mbox{if $j$ odd}
\end{array} \right.
\end{equation}
and assuming that the positions $j_1\ldots j_n$ are
are chosen in increasing order these correlation functions are given by
\begin{equation}
\langle \eta_{j_1}\eta_{j_2}\ldots\eta_{j_n}\rangle \;=\;
\frac{1}{Z}\, \langle W|
C^{j_1-1}\, F_{j_1}\, C^{j_2-j_1-1} \,F_{j_2} \,C^{j_3-j_2-1}
\ldots C^{j_n-j_{n-1}-1} \,F_{j_n}\, C^{2N-j_n} |V \rangle\,.
\end{equation}
Using the representations (\ref{Rep11})-(\ref{Norm2})
it is easy to check that
\begin{equation}
F_j C^{k-j-1} F_k \;=\; F_j C^{k-j}
\hspace{20mm}
(j<k)
\end{equation}
so that the $n$-point correlation functions reduce to the
one-point function $\langle \eta_{j} \rangle$:
\begin{equation}
\label{Reduction}
\langle \eta_{j_1}\eta_{j_2}\ldots\eta_{j_n}\rangle \;=\;
\langle \eta_{j_1} \rangle \;=\;
Z^{-1}\, \langle W|
C^{j_1-1}\, F_{j_1}\, C^{2N-j_1} |V \rangle\,.
\end{equation}
For $\alpha \neq \beta$ the one-point function reads
\begin{eqnarray}
\label{OnePointNeq}
\langle \eta_{2j} \rangle &=&
\frac{1}{Z_1} \, \alpha^{2N+1-2j} \,
(1-\beta) \, (\alpha^{2j}-\beta^{2j}) \\
\langle \eta_{2j-1} \rangle &=&
\frac{1}{Z_1} \, \alpha^{2N+2-2j} \,
\Bigl(\alpha^{2j-1}(1-\beta)-\beta^{2j-1}(1-\alpha)\Bigr)
\end{eqnarray}
whereas at the transition line $\alpha=\beta$ we have
\begin{eqnarray}
\label{OnePointEq}
\langle \eta_{2j} \rangle &=&
\frac{1}{Z_2} \, 2j\,(1-\alpha) \\
\langle \eta_{2j-1} \rangle &=&
\frac{1}{Z_2} \, \Bigl(\alpha+(2j-1)(1-\alpha)\Bigr)\,.
\end{eqnarray}
Although we used the two-dimensional matrices at this point, Eqs.
(\ref{Reduction})-(\ref{OnePointEq}) do not depend on the choice
of the representation since we have shown that the expectation
value of any sequence of operators is uniquely given
by the commutation relations of the algebra.
\\
\indent
Resubstituting $\tau_j$ into Eq. (\ref{Reduction}) we obtain an exact
expression for the $n$-point density correlation functions
$\langle \tau_{j_1} \tau_{j_2} \ldots  \tau_{j_n} \rangle$.
Denoting $\sigma_j = j\;mod\;2$, they read
\begin{equation}
\langle \tau_{j_1} \tau_{j_2} \ldots  \tau_{j_n} \rangle \;=\;
\alpha^n \prod_{i=1}^n \sigma_i \,+\,
\sum_{k=1}^n \Bigl( \prod_{i=1}^ {k-1} \sigma_l \Bigr)
\alpha^{k-1} (1 + \alpha \sigma_k - \beta - \beta\sigma_k)
\, \langle \eta_{j_k} \rangle \,.
\end{equation}
As a special case this formula includes the two-point correlation
functions which have been given as conjectures in Ref. \cite{Gunter}.

The example of the asymmetric exclusion model shows that the powerful
matrix product formalism can be applied successfully to models with
parallel dynamics. Since models of this type are widely studied, it
would be interesting to find further examples in order to understand
under which conditions the matrix product technique can be applied.
In particular it would be interesting solve the same model on a ring in
the presence of a defect. From this one could learn how to solve the full
exclusion process (with stochastic hopping in both directions) on
a ring with a defect \cite{RingDefect}. Despite of intensive efforts
the exact solution to this problem is not yet known.
%
%
%
\newpage
\noindent
{\bf Acknowledgements}\\[2mm]
I would like to thank E. Domany, I. Peschel, V. Rittenberg, S. Sandow and
G.M. Sch\"utz for helpful hints and fruitful discussions. I would also like
to thank the Minerva foundation for financial support.
%
%
%
%
%

\end{document}